\documentclass[conference]{IEEEtran}
\IEEEoverridecommandlockouts
\usepackage{cite}
\usepackage{amsmath,amssymb,amsfonts}
\usepackage{algorithmic}
\usepackage{graphicx}
\usepackage{subcaption}
\usepackage{textcomp}
\usepackage{xcolor}
\usepackage{stfloats}
\usepackage{flushend}
\def\BibTeX{{\rm B\kern-.05em{\sc i\kern-.025em b}\kern-.08em
    T\kern-.1667em\lower.7ex\hbox{E}\kern-.125emX}}
\begin{document}

\title{A Comprehensive PPG-based Dataset for HR/HRV Studies}

\author{
\IEEEauthorblockN{Jingye Xu\IEEEauthorrefmark{1}, Yuntong Zhang\IEEEauthorrefmark{2}, Wei Wang\IEEEauthorrefmark{1}, Mimi Xie\IEEEauthorrefmark{1}, Dakai Zhu\IEEEauthorrefmark{1}}
\IEEEauthorblockA{
\{jingye.xu, wei.wang, mimi.xie, dakai.zhu\}@utsa.edu\\
yuzhang@pvamu.edu\\
\IEEEauthorrefmark{1}\textit{the University of Texas at San Antonio, TX, USA}\\
\IEEEauthorrefmark{2}\textit{Prairie View A\&M University, TX, USA}}\\
}
\maketitle

\begin{abstract}
Heart rate (HR) and heart rate variability (HRV) are important vital signs for human physical and mental health. Recent research has demonstrated that photoplethysmography (PPG) sensors can infer HR and HRV. However, it is difficult to find a comprehensive PPG-based dataset for HR/HRV studies, especially for various study needs: multiple scenes, long-term monitoring, and multimodality (multiple PPG channels and extra acceleration data).
In this study, we collected a comprehensive multimodal long-term dataset to address the gap of missing an all-in-one HR/HRV dataset (denoted as UTSA-PPG).
We began by reviewing state-of-the-art datasets, emphasizing their strengths and limitations. Following this, we developed a custom data acquisition system and then collected the UTSA-PPG dataset and compared its key features with those of existing datasets. Additionally, five case studies were conducted, including comparisons with state-of-the-art datasets. The outcomes highlight the value of our dataset, demonstrating its utility for HR/HRV estimation exploration and its potential to aid researchers in creating generalized models for targeted research challenges.
\end{abstract}


%%%%%%%%%%%%%%%%%%%%%%%%%%%%%%%%%%%%%%%%%%%%%%%%%%%%%%%%%
\section{Introduction}
\label{introduction}

Heart rate (HR) and heart rate variability (HRV) are important vital signs for the cardiovascular system and have been widely used as biomarkers for diagnostic and early prognostic of several diseases such as hypertension and heart failure~\cite{fox2007resting,2012-Xhyheri-HRVToday,acharya2006heart}. Therefore, it is desirable to have a real-time HR/HRV monitoring system, which can conveniently provide accurate data in an effective manner to support health monitoring and such applications.
The traditional and reliable approach to continuously monitoring HR/HRV is to utilize electrocardiogram (ECG) devices, which record the heart's rhythm. However, ECGs can be expensive and require attaching several electrodes to human bodies, which can be inconvenient to deploy for outpatients or other users to operate in a continuous manner. 
As an alternative to ECG, photoplethysmography (PPG) sensors, which monitor light signal changes in blood flows, can also be exploited to derive HR/HRV~\cite{2007-Wang-TBCS-PPGEar,zhang2014troika}.
A PPG sensor can be placed on the surface of human skin, like fingers, wrist, or earlobe, to provide HR/HRV readings and is more convenient to use.
Given their low cost and convenience, PPG sensors have been widely utilized as an inexpensive alternative to monitor HR/HRV in wearable embedded devices (such as smartwatches), which have limited energy and resources~\cite{bhowmik2017novel,phan2015smartwatch}.
It is worth noting that PPG signals are highly susceptible to motion artifacts (MA), prompting most existing research to prioritize noise removal for accurate HR/HRV estimation using either signal processing methods or machine learning techniques. Regardless of the chosen approach, dataset remains a critical factor in these studies. Consequently, numerous datasets have been collected in the HR/HRV estimation field to address this demand. However, these datasets have several limitations, which we have summarized below:

\begin{enumerate}
    \item Single experimental scene\cite{BIDMC-dataset,mimicperform-dataset,liang2018new-ppg-bp-dataset, goldberger2000physiobank,but-ppg-original}. Some datasets are limited to a single scenario, which significantly restricts their ability to capture the diversity of real-world conditions. In such cases, the data may lack representation of various motion artifacts, such as those caused by sudden accelerations (Acc), rotations, or changes in perspective. These motion artifacts are often critical for evaluating the robustness and adaptability of algorithms in dynamic environments both in signal processing and machine learning methods. Consequently, datasets with only one scenario may fail to provide sufficient challenges for comprehensive testing and development, limiting their applicability to broader, more complex use cases.
    \item Short session length\cite{but-ppg-original,zhang2014troika, nemcova2021brno-but-ppg-version2, liang2018new-ppg-bp-dataset}. Some datasets are characterized by very short data collection durations, which makes them unsuitable for HRV studies and long-term monitoring applications.
    % Calculating HRV typically requires at least 30 seconds of continuous data~\cite{acharya2006heart}. But datasets with shorter durations fail to meet this basic requirement, significantly limiting their applicability.
    Besides, short session lengths lack sufficient historical information, which is critical for analyzing trends, detecting patterns, and making predictions over extended periods.
    As a result, such datasets are constrained to specific, limited applications and cannot support comprehensive studies that demand extended observation and robust historical context.
    \item Lack of multimodality\cite{BIDMC-dataset, mimicperform-dataset, liang2018new-ppg-bp-dataset,goldberger2000physiobank, but-ppg-original, nemcova2021brno-but-ppg-version2}. Some datasets do not have extra PPG channels or accelerometer data, limiting their ability to effectively address complex research questions. For instance, MA is a major factor influencing the accuracy of HR/HRV estimation. Incorporating accelerometer data is crucial for detecting and compensating for these artifacts, improving the reliability of the estimation. Furthermore, providing multiple modalities-such as combining PPG channel data (red, green, and infrared light signals)-enables a deeper understanding of how each modality impacts HR/HRV estimation performance. Multimodal datasets allow researchers to explore the complementary strengths of different signals and develop more robust, adaptable algorithms for real-world applications. Without such diversity, datasets are less effective for advancing comprehensive, multimodal approaches to HR/HRV estimation.
\end{enumerate}

To address the aforementioned limitations, we collected the UTSA-PPG dataset\footnote{https://github.com/utsanpb2-234/UTSA-PPG}, which was designed to overcome the constraints of existing datasets. the UTSA-PPG dataset features a diverse range of scenarios, capturing various MA to ensure robust algorithm evaluation in dynamic and real-world environments. Besides, it includes extended data collection durations, enabling both short-term and long-term analysis, making it capable of HR/HRV estimation over sufficient time windows. Additionally, the UTSA-PPG dataset embraces multimodality, incorporating signals such as acceleration data alongside multiple channels of PPG signals. This enables comprehensive studies on the impact of each modality and supports the development of advanced, multimodal solutions for HR and HRV estimation. By addressing these critical gaps, the UTSA-PPG dataset provides a versatile and reliable foundation for HR/HRV research and innovation. The main contributions are as follows:

\begin{enumerate}
    \item We conducted a thorough review of existing state-of-the-art HR/HRV datasets, analyzing their strengths and limitations (Section~\ref{sec:related_works}).
    \item A custom PPG data acquisition system tailored to address identified gaps, and the UTSA-PPG dataset, which contains multiple scenarios, is multimodal, and accommodates both short-term and long-term HR/HRV studies, was collected (Section~\ref{sec:dataset}).
    \item Five case studies were conducted to compare the UTSA-PPG dataset and state-of-the-art datasets. The results demonstrated the practicality and advantages of the UTSA-PPG dataset over existing datasets (Section~\ref{sec:case_study}).
\end{enumerate}

\section{Related Works}
\label{sec:related_works}

\begin{table*}[t]
  \centering
  \caption{Datasets for HR/HRV estimation.}
  \resizebox{\textwidth}{!}{
  \begin{tabular}{lcccccccccc}
    \hline
    Dataset & \# Sub & \# Sessions & Session Length & Sampling Rate & PPG Channels & Other Features & Time Window & Placements & Ground Truth & Year \\
    \hline
    \multicolumn{11}{c}{Datasets from hospitals} \\
    \hline
    BIDMC~\cite{BIDMC-dataset} & - & 53 & 8min & 125Hz & 1 & - & Adaptive & Finger & ECG(2-lead) & 2017 \\
    MIMIC PERform Train~\cite{mimicperform-dataset} & 200 & 200 & 10min & 125Hz & 1 & - & Adaptive & Finger & ECG(2-lead) & 2022 \\
    MIMIC PERform Test~\cite{mimicperform-dataset} & 200 & 200 & 10min & 125Hz & 1 & - & Adaptive & Finger & ECG(2-lead) & 2022 \\
    PPG-BP~\cite{liang2018new-ppg-bp-dataset} & 219 & 657 & 2.1s & 1kHz & 1 & - & 2.1s & Finger & - & 2018 \\
    \hline
    \multicolumn{11}{c}{Datasets from laboratories} \\
    \hline
    PPG-DaLiA~\cite{reiss2019deep-ppg-dalia} & 15 & 15 & 2.5h & 64Hz & 1 & Acc(32Hz,700Hz) & 8s & Wrist & ECG(3-lead) & 2019 \\
    WESAD~\cite{schmidt2018introducing-wesad} & 15 & 15 & 1.5h & 64Hz & 1 & Acc(32Hz,700Hz) & Adaptive & Wrist & ECG(3-lead) & 2018 \\
    ISPC~\cite{zhang2014troika} & 12 & 12 & 5min & 125Hz & 2 & Acc(125Hz) & 8s & Wrist & ECG(chest band) & 2014 \\
    BUT PPG~\cite{but-ppg-original,nemcova2021brno-but-ppg-version2} & 50 & 3,888 & 10 s & 30Hz & 1 & Acc(100Hz) & 10s & Finger, Ear & ECG (1-lead) & 2024 \\
    Rice iPPG~\cite{pai2021hrvcam-rice-ippg-dataset} & 14 & 70 & 2min & 30Hz & 1 & - & 2min & Face(Remote) & Pulse Oximeter & 2021 \\
    UTSA-PPG (this work) & 12 & 36 & 30min & 100Hz & 3 & Acc(100Hz) & Adaptive & Wrist, Finger & ECG(3-lead) & 2025 \\
    \hline
  \end{tabular}}
  \label{tab:datasets}
\end{table*}

Datasets used for HR/HRV estimation can be broadly categorized into two types based on their sources: those collected from hospitals and those gathered in laboratories. However, these two types differ in their primary focus.

Hospital datasets primarily aim to establish connections between PPG signals and patients’ overall health status. As a result, these datasets typically contain fewer MA and offer high-quality signal waveforms.
The MIMIC Clinical Database series~\cite{saeed2011multiparameter-mimic-II, johnson2016mimic-mimic-III, goldberger2000physiobank} include data recorded from real Intensive Care Unit (ICU) patients. The datasets contain many annotations regarding different medical needs: ECG beat labels, patient status alarms, pulmonary arterial pressure, central venous pressure, PPG signals, et al, making it possible to predict various human health indicators and statuses. Because the original MIMIC Clinical Database is huge and not the whole dataset is suitable for HR/HRV estimation, some derivative datasets were extracted to focus on HR/HRV related fields.
The BIDMC Dataset is a subset of Physionet’s MIMIC II Waveform Database Matched Subset~\cite{goldberger2000physiobank,saeed2011multiparameter-mimic-II}, which contains all MIMIC II Waveform Database records that have been associated with the MIMIC II Clinical Database records. Although it is intended to evaluate the performance of respiratory rate algorithms, the dataset contains both PPG and ECG signals, making it feasible for HR/HRV estimation as well. 
Beside, there are another two datasets which are named MIMIC PERform Train Dataset and MIMIC PERform Test Dataset~\cite{mimicperform-dataset}, respectively. Different from the BIDMC Dataset, MIMIC PERform Datasets are extracted from the MIMIC III Database~\cite{johnson2016mimic-mimic-III,goldberger2000physiobank}, and the duration is around 10min. Furthermore, MIMIC PERform datasets have 200 subjects.
Besides the MIMIC series and its derivatives, there is another dataset, the PPG-BP~\cite{liang2018new-ppg-bp-dataset}, was collected from hospital and intended to explore and improve the understanding of relationships between cardiovascular health and PPG signals. With ECG as the ground truth, PPG-BP can be used to study HR estimation as well. However, PPG-BP could not be the perfect dataset to evaluate the HRV estimation due to its short session length.

In contrast to hospital datasets, laboratory datasets focus on accurately calculating HR/HRV in complex environments. Consequently, these datasets often include MA and encompass a variety of scenarios.
PPG-DaLiA~\cite{reiss2019deep-ppg-dalia} dataset comprises activities of daily living, including: sitting, working, walking, driving, playing games, et al. As daily activities involve many MA, PPG-DaLiA provides a good chance to analyze how to eliminate MA and improve the PPG HR/HRV estimation.
WESAD~\cite{schmidt2018introducing-wesad} dataset is originally used to induce different emotions: baseline, stress, amusement, and meditation. With the same hardware equipments with PPG-DaLiA, WESAD dataset could be used for HR/HRV estimation research, too.
ISPC~\cite{zhang2014troika} dataset is specifically used to study how to eliminate MA' effect on HR estimation. All data in ISPC are sampled at 125Hz and calculated HRs from ECG signals are provided directly. The ISPC dataset is good for HR estimation; however, due to the short session length, it does not fit long-term application, especially HRV estimation.
BUT PPG~\cite{but-ppg-original,nemcova2021brno-but-ppg-version2} is a dataset comprising 3,888 10-second recordings of PPGs and could be used for the research of PPG quality evaluation and estimation. PPG signals are collected on finger from the rear camera or ear from the front camera. BUT PPG dataset can be used to analyze the feasibility and effectiveness of HR estimation with smartphones due to its distinctive PPG data.
Besides, Rice iPPG~\cite{pai2021hrvcam-rice-ippg-dataset} dataset collects imaging photoplethysmography signals acquired using cameras and provides researchers the resources to investigate non-contact iPPG based HRV and HR measurements. The dataset consists of facial video recordings of 14 subjects under different engagement activities such as reading, talking, watching, and deep breathing. It also contains videos when they were stationary to establish a baseline.

Table~\ref{tab:datasets} displays the main characteristics of the aforementioned HR/HRV estimation datasets collected from hospitals and laboratories. It is worth noting that hospital datasets typically place PPG sensors on patients’ fingers and lack multiple PPG channels. This preference can be attributed to the fact that finger-mounted PPG sensors are more effective in detecting blood vessel changes in the fingers. Additionally, hospital datasets often include only a single PPG channel because adding extra channels would introduce unnecessary inconvenience for patients. Furthermore, hospital datasets do not monitor movement features, as they assume patient movement is minimal. If movement affects the PPG signals, such data is typically discarded. Unlike hospital datasets, laboratory datasets primarily focus on accurately estimating HR/HRV from less pristine PPG signals (affected by environmental noises). As a result, these datasets typically include acceleration data to account for MA and often feature multiple scenarios for enhanced analysis. In terms of PPG sensor placement, laboratory datasets offer more flexibility, exploring options beyond simply mounting the sensor on fingers.

By summarizing the aforementioned datasets, we observe that existing datasets often meet specific needs but fail to address all requirements comprehensively. Some datasets are limited to a single scenario, such as hospital datasets, which restrict their ability to capture the diversity of real-world conditions. Others, like PPG-BP, BUT PPG, and ISPC, have short data session lengths, making them unsuitable for long-term HRV estimation. Additionally, certain datasets, such as hospital datasets and BUT PPG, lack multimodality, limiting their effectiveness in addressing complex research questions. Therefore, there is a need for a comprehensive, multiple scenarios, multimodal, long-term PPG dataset. Thus, we collected the UTSA-PPG dataset, with its main characteristics outlined in Table~\ref{tab:datasets}. The UTSA-PPG dataset includes 12 subjects, with each participant taking part in three recording sessions, covering three distinct activity scenarios. Each session lasted at least 30 minutes to support long-term studies. Additionally, the dataset contains three channels of PPG signals along with Acc data, providing a platform to explore the impact of different modalities on HR/HRV research. Furthermore, UTSA-PPG dataset also provides PPG data from two commercial devices for cross-comparison. These features make it an all-in-one resource for comprehensive HR/HRV studies.
\section{Dataset}
\label{sec:dataset}

\subsection{Data Collection Setup}

\begin{figure}
    \centering
    \includegraphics[width=0.6\linewidth]{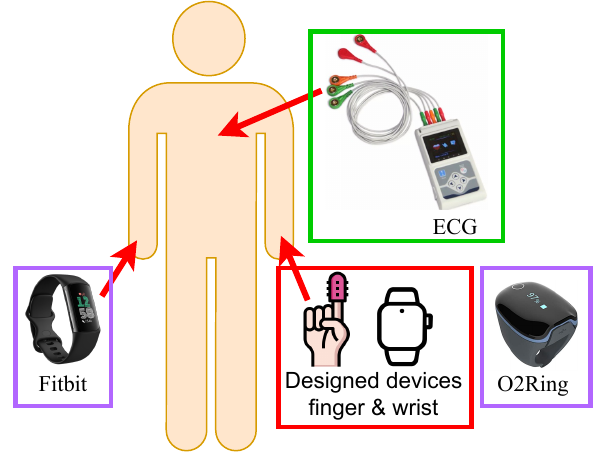}
    \caption{Data collection setup.}
    \label{fig:setup}
\end{figure}

\begin{figure}
    \centering
    \includegraphics[width=0.6\linewidth]{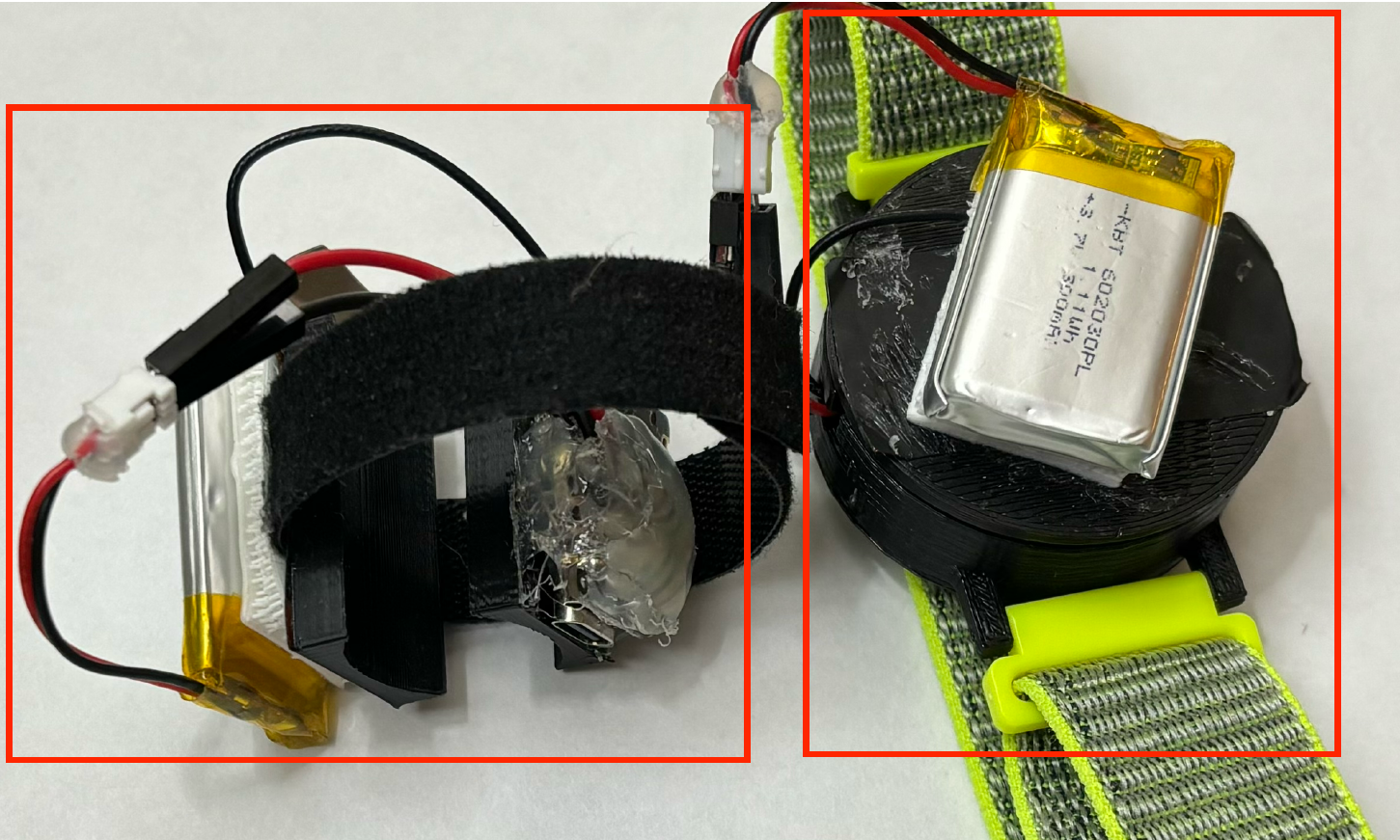}
    \caption{PPG devices for finger and wrist.}
    \label{fig:hardware}
\end{figure}

As shown in Fig.~\ref{fig:setup}, subjects wear the data acquisition system on their fingers, wrists, and torso.
The ECG (green box) was connected to the human torso using cables and electrodes.
Additionally, two customized PPG data collection devices (red box) worked simultaneously: a finger clip and a wristband, as illustrated in Fig.~\ref{fig:hardware}.
The finger clip was mounted on the subject's left middle fingertip, and the wristband was installed on the subject's left wrist.
The subjects also had two commercial devices (purple boxes) worn simultaneously, a smart ring on the left thumb and a smartwatch on the right wrist. These settings ensure that our dataset supports multimodal research.

\begin{figure}
    \centering
    \includegraphics[width=0.6\linewidth]{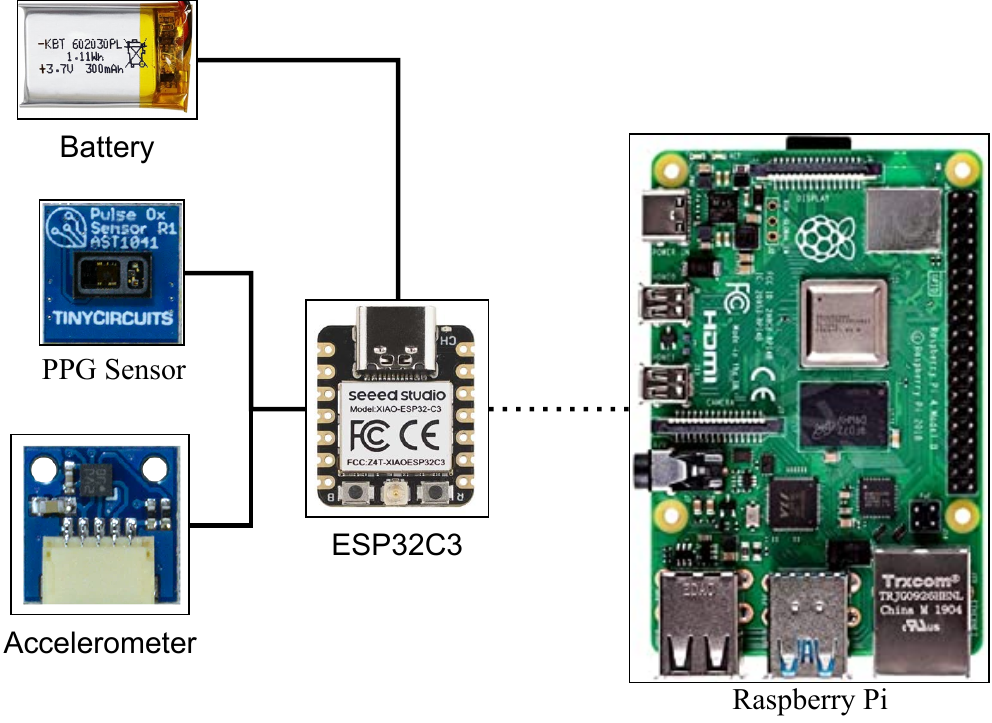}
    \caption{PPG devices connection.}
    \label{fig:hw_connect}
\end{figure}

Our data acquisition system contains:

\begin{enumerate}
    \item Ground Truth: A TLC5007 Dynamic 3-lead ECG holter\footnote{https://contechealth.com/collections/ecg-eeg-machine-holter/products/tlc9803-dynamic-ecg-systems} requires five electrodes to be attached to the torso during data collection. It records the heart activity and stores it on the device.
    \item PPG Devices: Two PPG sensors, AST1041, are used to collect data on fingertip and wrist simultaneously\footnote{https://tinycircuits.com/products/pulse-oximetry-sensor-wireling} as shown in Fig.~\ref{fig:hardware}. The left red rectangle shows a clip that goes around a finger, which contains a PPG sensor to record infrared (IR) and red light signals. The right red rectangle shows a wristband that contains a PPG sensor to record the green light signals. In addition, an accelerometer, AST1001 \footnote{https://tinycircuits.com/products/accelerometer-wireling-bma250}, is added for our wristband.
    \item PPG Controllers: Two microcontrollers, XIAO ESP32C3\footnote{https://www.seeedstudio.com/Seeed-XIAO-ESP32C3-p-5431.html} (32 bit, 160 MHz, 400 KB SRAM, and 4 MB flash memory) are integrated into the finger and wrist devices. These microcontrollers collect data from the PPG sensors and accelerometers locally and transmit it to the Raspberry Pi via Bluetooth, as depicted in Fig.~\ref{fig:hw_connect}. This setup enables subjects to move more freely during data collection.
    \item Commercial Devices: Two commercial devices are selected to collect data simultaneously, as shown in Fig.~\ref{fig:setup}. One is a smartwatch, and the other one is a smart ring.
    \begin{enumerate}
        \item Fitbit Charge 6\footnote{https://www.fitbit.com/global/us/products/trackers/charge6} (smartwatch). 
        \item O2Ring Continuous Oxygen Monitor\footnote{https://getwellue.com/products/o2ring-wearable-pulse-oximeter} (smart ring). 
    \end{enumerate}
    \item Power Supplies: The finger clip and wristband devices are powered by small lithium batteries (3.7V 300mAh), and a power cable powers the Raspberry Pi. Two commercial devices are powered by their built-in batteries. The ECG holter is powered by 2 AA batteries.
\end{enumerate}

\subsection{Data Collection Protocol}

Fifteen subjects participated in the study. However, three subjects' data were abandoned due to system failure on the ECG devices, making the UTSA-PPG dataset contain twelve subjects. Each subject followed a defined data collection protocol, including three different scenarios: sit, sleep, and office work. Each scenario was one data collection session that lasted for a minimum of 30 minutes and was collected separately, giving subjects time to rest during every two sessions. Before each data collection session, the timestamps of all devices were checked and synced to ensure that the time difference was as low as possible. When collecting data, subjects were required to follow the rules below for different scenarios:

\begin{itemize}
    \item Sit: The subject sits on a chair, putting hands on the desk, with actions such as working with a laptop/smartphone, talking to people, and drinking water.
    \item Sleep: The subject relaxes and attempts to take a nap in bed.
    \item Office work: Besides the activities in the Sit scenario, the subject involves a walking activity.
\end{itemize}

\subsection{Data Processing}

\subsubsection{ECG Data}

After data collection was completed, the ECG data was exported to a computer and analyzed by the software that comes with the device. The software analyzed the recorded ECG data, displayed the electrocardiogram, and exported the RR intervals. Then, the ground truth HRs/HRVs were calculated using the exported RR intervals and synchronized with PPG data through timestamps.

\subsubsection{PPG and Acc Data}

PPG and Acc data were collected at 100Hz. The finger PPG data contains red and IR light signals and wrist PPG data contains green light signals. We found that different subjects and different light signals have different reading ranges, which may come from personal biases such as skin tone. Therefore, a bandpass filter was applied to keep signals between 0.5Hz and 4Hz to remove such noises. Moreover, the timestamps were used to sync PPG and Acc data, as well.

\subsubsection{Commercial Device Data}

Both commercial devices come with software that can display and export monitoring results. However, they only export HR data but not HRV data. Fitbit provides HRV trends only during sleep, and O2Ring does not offer HRV information at all. Therefore, only the HR monitoring results were extracted from these devices. Fitbit provided around one HR reading every two seconds, and O2Ring provided one HR reading every four seconds.

\subsection{Ethics}

The collection of the UTSA-PPG dataset was approved by the Institutional Review Board (IRB) to ensure that ethical standards were maintained throughout the process. Besides, each subject was fully informed about the procedure of dataset collection and the purpose of the dataset. All participants signed the written consent before their involvement in the study. Furthermore, to protect the privacy of the individuals, all subjects in the dataset were de-identified, ensuring that their personal information remained confidential.

\section{Experiments and Case Studies}
\label{sec:case_study}

In this section, five case studies were conducted to explore the potential of the UTSA-PPG dataset in HR/HRV estimation. For HR estimation, we examined the dataset’s generalization capabilities (Section ~\ref{sec:compare_datasets}), the impact of different PPG channels on HR estimation accuracy (Section ~\ref{sec:compare_channel}), and performance across various scenarios (Section ~\ref{sec:compare_scenario}). For HRV estimation, we investigated the dataset’s ability to handle different data history lengths for diverse HRV monitoring needs (Section ~\ref{sec:compare_data_length}) and the optimal mode selection for specific HRV monitoring applications (Section ~\ref{sec:compare_model}).

\subsection{HR estimation}

\subsubsection{HR estimation with different datasets}
\label{sec:compare_datasets}

\begin{figure}
    \centering
    \includegraphics[width=0.9\linewidth]{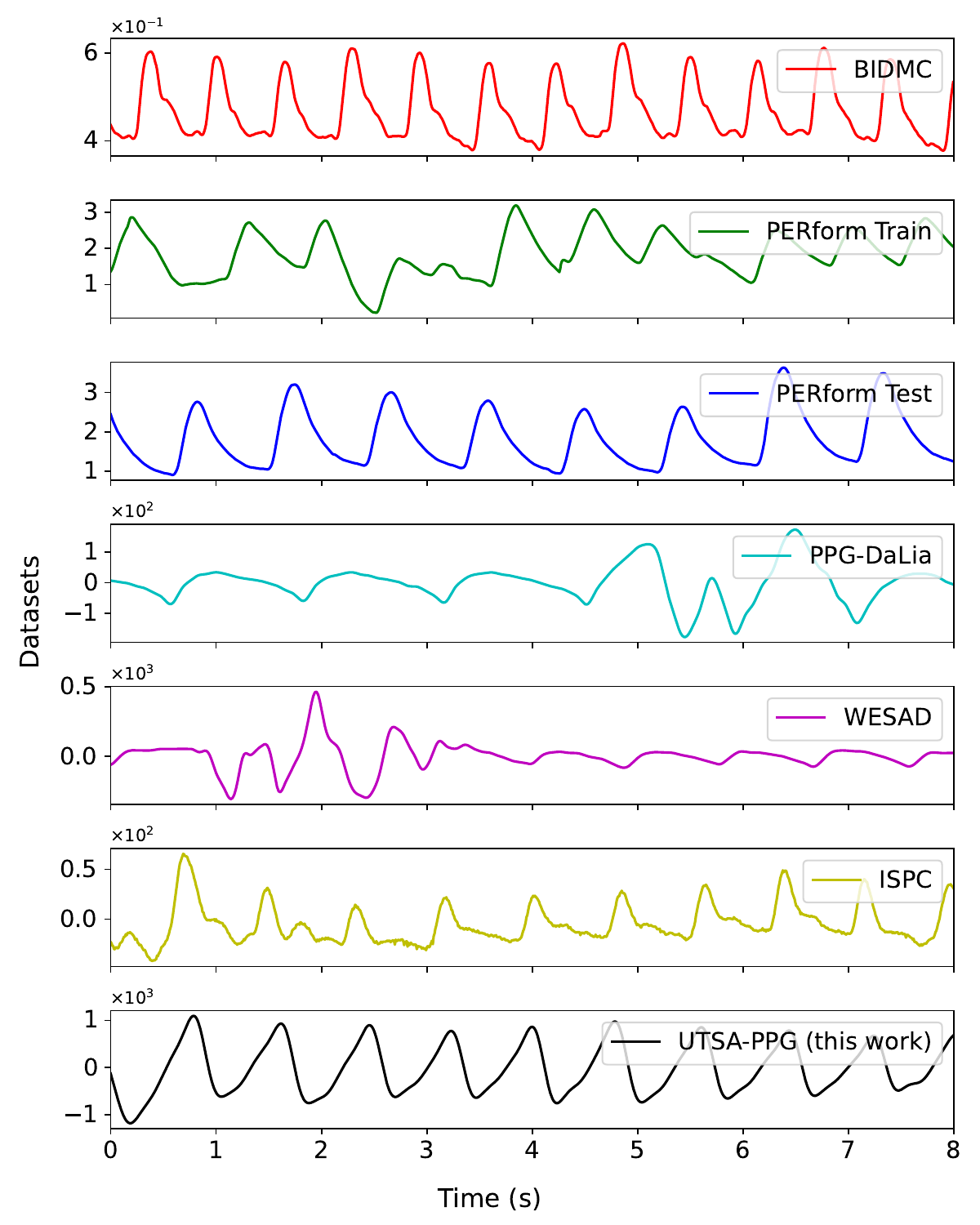}
    \caption{Preview of different datasets' segments with a window size of 8s.}
    \label{fig:preview}
\end{figure}

% how to choose datasets
We first explored the HR estimation with different datasets to examine the generalization capability. All datasets in Table~\ref{tab:datasets} were used except for PPG-BP, BUT PPG, and Rice iPPG. PPG-BP has a very short session length; BUT PPG and Rice iPPG's data are both from cameras, making them infeasible to compare with others. When processing the PPG data of all datasets, a sliding window of 8s was chosen without overlap so that every dataset can maintain the same format. For each PPG waveform, the ground truth (ECG-derived HR) was directly provided from the dataset if available, or calculated from the provided ECG waveform. WFDB~\cite{xie2021waveform_wfdb,goldberger2000physiobank} was used when we need to calculate the HR from ECG waveform.
Fig.~\ref{fig:preview} displays different datasets' segments with a window size of 8s after processing.

% evaluate methods
Two main methods were employed: signal processing (SP) and machine learning (ML).
When utilizing signal processing solely to conduct HR estimation, the PPG data was directly sent to the HeartPy~\cite{van2019heartpy} toolkit to calculate the HR. Then, the Mean Absolute Error (MAE) and Mean Absolute Percentage Error (MAPE) were calculated based on ground truth. The results of HR estimation with signal processing are shown in Table~\ref{tab:sp}. It can be found that the BIDMC dataset exhibits lowest MAE and MAPE, attributed to its high-quality data. However, despite also being hospital-based datasets, the MIMIC PERform Train and Test datasets show poor HR estimation performance when using HeartPy. This is due to the presence of MA and periods of low-quality data within these datasets~\cite{mimicperform-dataset}. Additionally, PPG-DaLia and WESAD show the highest MAE and MAPE, as their data collection involves numerous real-life MA, making it challenging to calculate HR purely through signal processing. In contrast, ISPC and UTSA-PPG datasets demonstrate relatively low MAE and MAPE, indicating that both datasets offer high-quality data while still including some MA.

\begin{table}
    \centering
    \caption{HR estimation with SP method: HeartPy.}
    \begin{tabular}{lccc}
    \hline
    Dataset & \# Samples & MAE & MAPE \\
    \hline
    \multicolumn{4}{c}{Datasets from hospitals} \\
    \hline
    BIDMC               &  3,180   & 1.735  & 0.021 \\
    MIMIC PERform Train &  15,000  & 16.203 & 0.191 \\
    MIMIC PERform Test  &  15,000  & 17.813 & 0.190 \\
    \hline
    \multicolumn{4}{c}{Datasets from laboratories} \\
    \hline
    PPG-DaLiA           &  16,175  & 18.347 & 0.223 \\
    WESAD               &  10,849  & 20.951 & 0.299 \\
    ISPC                &  444     & 14.466 & 0.112 \\
    UTSA-PPG (this work)&  10,568  & 6.952  & 0.091 \\
    \hline
    \end{tabular}
    \label{tab:sp}
\end{table}

When using machine learning exclusively to predict HR, the PPG waveforms served as the features, while the ECG-derived HRs were used as the labels. Initially, all signals were upsampled to 500Hz using SciPy’s resample function\footnote{https://docs.scipy.org/doc/scipy-1.15.0/index.html} to ensure compatibility for cross-validation with the trained machine learning models. Next, each dataset was normalized using z-scores and then split into two parts: 80\% for training and 20\% for testing. To evaluate HR estimation performance, three basic neural networks — convolutional neural network (CNN), gated recurrent units (GRU), and long short-term memory network (LSTM) — were employed. Simple network architectures were implemented rather than conducting network architecture searches (NAS), as the focus was on assessing the generalizability of models trained on different datasets. Tables~\ref{tab:cnn_structure},~\ref{tab:gru_structure}, and~\ref{tab:lstm_structure} give the details of our employed models.

\begin{table}
    \centering
    \caption{CNN model structure.}
    \begin{tabular}{llc}
    \hline
    Layer        & Output Shape         & \# Trainable Parameter \\
    \hline
    Conv1D       & (None, 3998, 512)    & 2,048       \\
    MaxPooling1D & (None, 1999, 512)    & 0           \\
    Conv1D       & (None, 1997, 256)    & 393,472     \\
    MaxPooling1D & (None, 998, 256)     & 0           \\
    Conv1D       & (None, 996, 128)     & 98,432      \\
    MaxPooling1D & (None, 498, 128)     & 0           \\
    Conv1D       & (None, 496, 64)      & 24,640      \\
    MaxPooling1D & (None, 248, 64)      & 0           \\
    Flatten      & (None, 15872)        & 0           \\
    Dense        & (None, 1024)         & 16,253,952  \\
    Dense        & (None, 512)          & 524,800     \\
    Dense        & (None, 256)          & 131,328     \\
    Dense        & (None, 1)            & 257         \\ 
    \hline
    \end{tabular}
    \label{tab:cnn_structure}
\end{table}

\begin{table}
    \centering
    \caption{GRU model structure.}
    \begin{tabular}{llc}
    \hline
    Layer        & Output Shape         & \# Trainable Parameter \\
    \hline
    GRU          & (None, 64)           &  12,864       \\
    Dense        & (None, 512)          &  33,280       \\
    Dense        & (None, 256)          & 131,328       \\
    Dense        & (None, 1)            & 257           \\
    \hline
    \end{tabular}
    \label{tab:gru_structure}
\end{table}

\begin{table}
    \centering
    \caption{LSTM model structure.}
    \begin{tabular}{llc}
    \hline
    Layer        & Output Shape         & \# Trainable Parameter \\
    \hline
    LSTM         & (None, 64)           &  16,896       \\
    Dense        & (None, 512)          &  33,280       \\
    Dense        & (None, 256)          & 131,328       \\
    Dense        & (None, 1)            & 257           \\
    \hline
    \end{tabular}
    \label{tab:lstm_structure}
\end{table}

\begin{figure}
    \centering
    \begin{subfigure}[b]{1\linewidth}
        \centering
        \includegraphics[width=\textwidth]{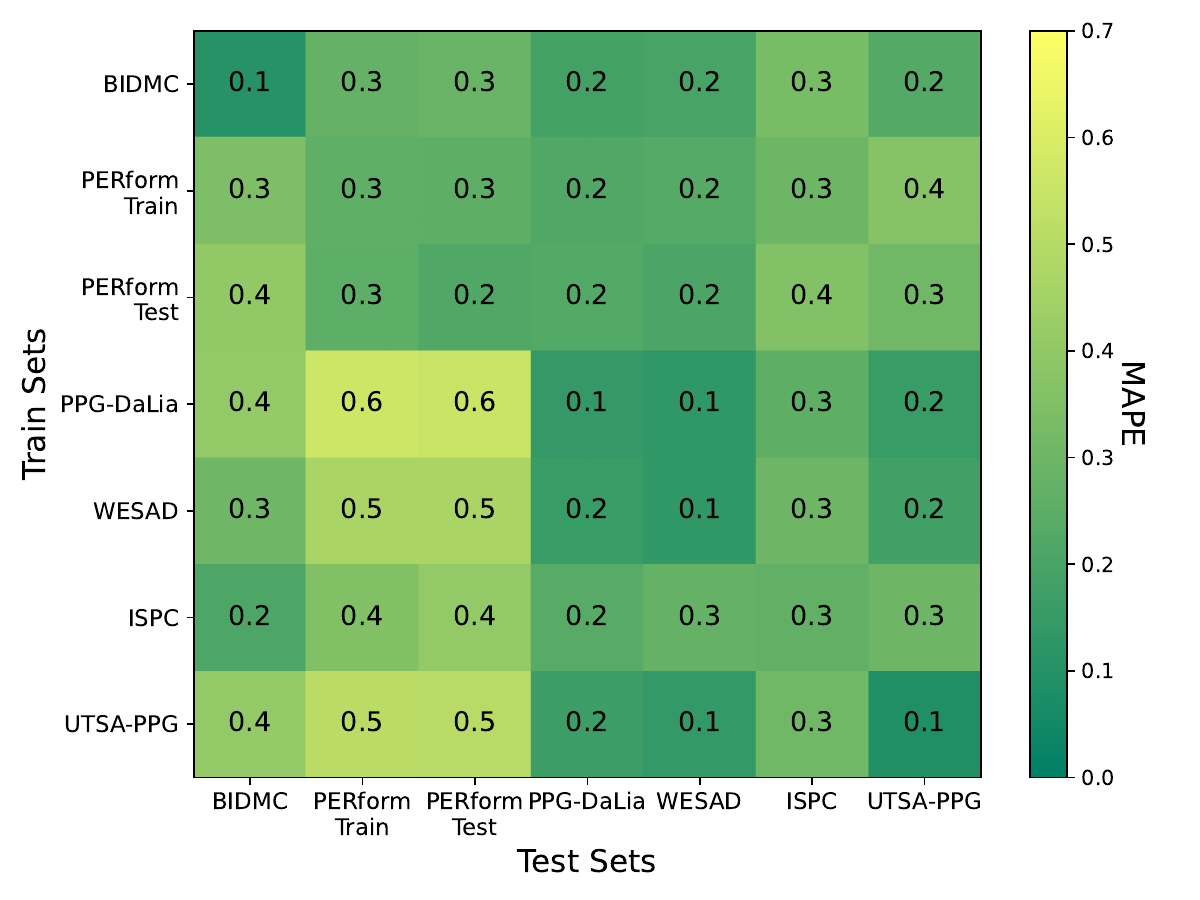}
        \caption{CNN.}
        \label{fig:cross_dataset_mape_cnn}
    \end{subfigure}
    
    \begin{subfigure}[b]{1\linewidth}
        \centering
        \includegraphics[width=\textwidth]{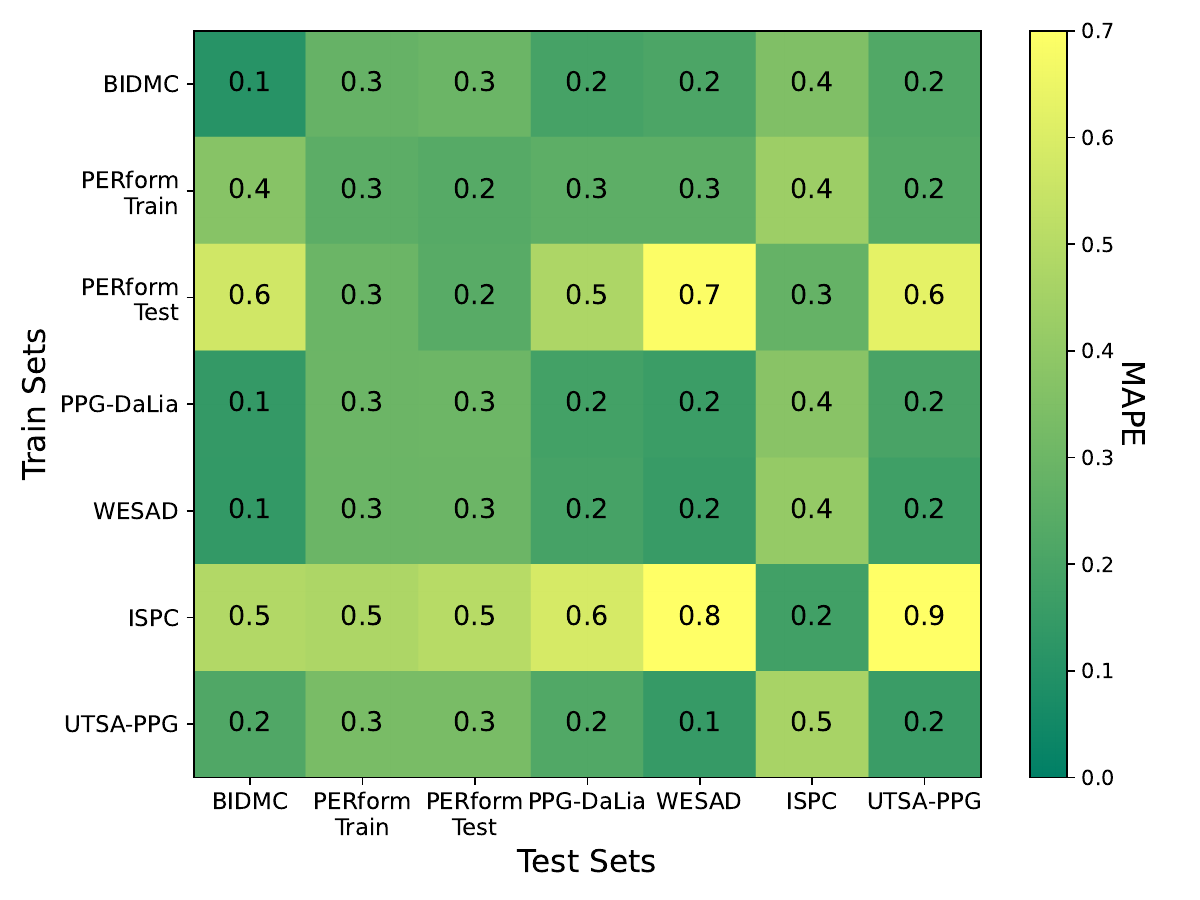}
        \caption{GRU.}
        \label{fig:cross_dataset_mape_gru}
    \end{subfigure}

    \begin{subfigure}[b]{1\linewidth}
        \centering
        \includegraphics[width=\textwidth]{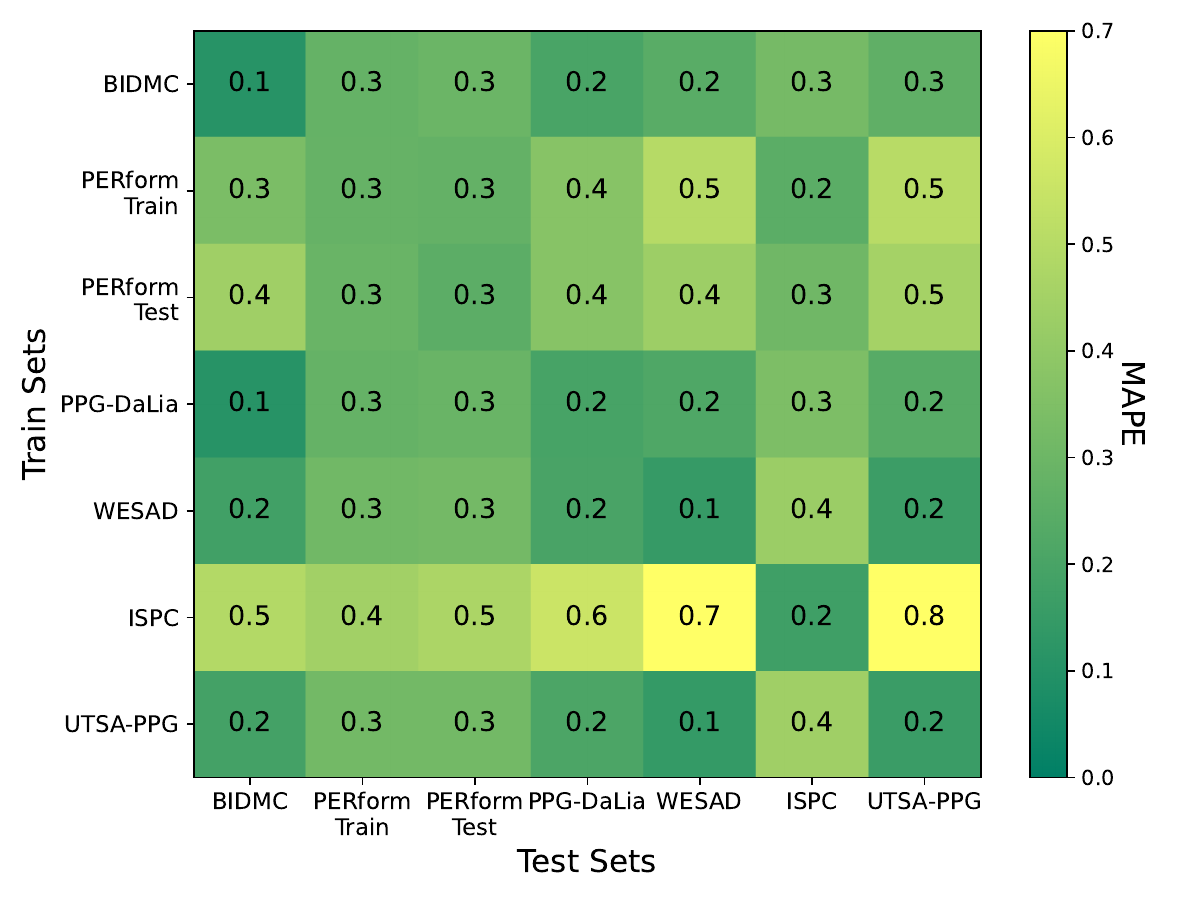}
        \caption{LSTM.}
        \label{fig:cross_dataset_mape_lstm}
    \end{subfigure}
    \caption{HR estimation with ML methods: CNN, GRU, LSTM.}
    \label{fig:ml}
\end{figure}

Fig.~\ref{fig:ml} shows the results of HR estimation with ML methods, where y-axis indicates which dataset's train set was used to train the model, and x-axis indicates which dataset's test set was used to test the model. Fig.~\ref{fig:cross_dataset_mape_cnn} displays the performance of training a simple CNN model using different datasets. It can be noticed that every dataset performs well when validating the model by its own test set. However, almost all models trained from laboratory datasets fail to recognize MIMIC PERform datasets, leading to very high MAPEs. It is worth noting that the performance gets better when training the models using GRU and LSTM, as shown in Fig.~\ref{fig:cross_dataset_mape_gru} and Fig.~\ref{fig:cross_dataset_mape_lstm}, respectively. Considering GRU and LSTM models perform well in time-series data, they are more suitable in estimating HR on a PPG waveform. Besides, all hospital models perform well when testing on hospital data because of the similar hospital scenario. For laboratory models, ISPC model could only recognize ISPC data, while performing bad on other datasets. The remaining laboratory models, PPG-DaLia, WESAD, and UTSA-PPG, show a similar good performance on all datasets, indicating the generalizability. Furthermore, all models show an MAPE above 0.1, that is to say, solely utilizing machine learning to predict HR is not the optimal approach. The conclusion here is consistent with Zhang~\cite{yuntong2022hrv}, who proposed a compound method that employs signal processing and machine learning together, which can drastically decrease the computing load while increasing the estimation accuracy. Consequently, to explore the UTSA-PPG datasets' potential in different aspects, we employed the compound method in the remaining four case studies.

\begin{figure*}[ht!]
    \centering
    \includegraphics[width=0.87\linewidth]{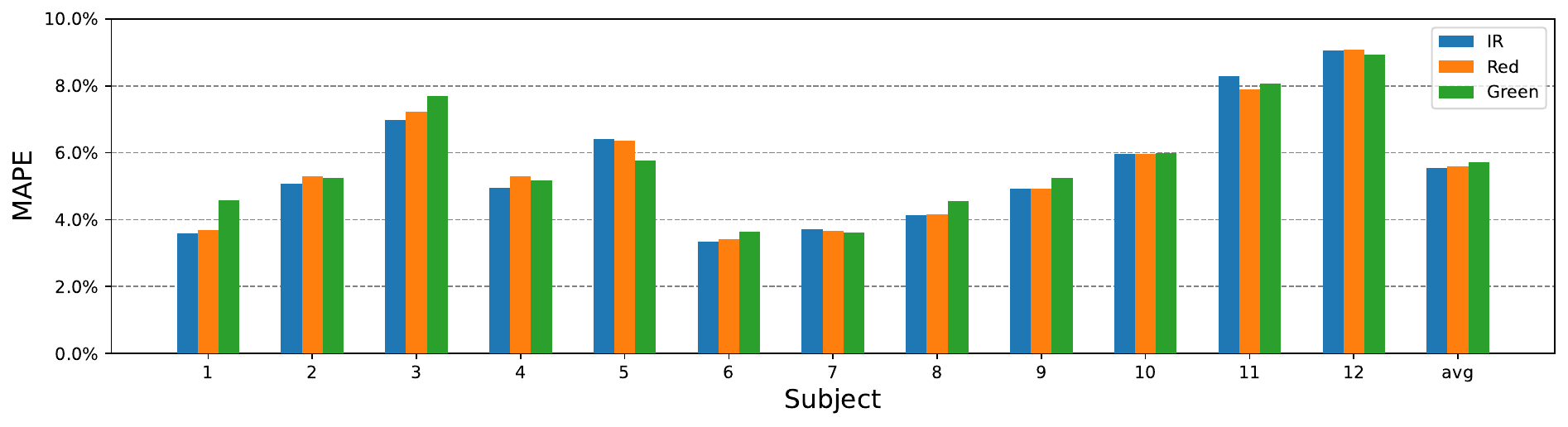}
    \caption{HR estimation with different channels in the sit scenario.}
    \label{fig:channel}
\end{figure*}

\subsubsection{HR estimation using different channels}
\label{sec:compare_channel}

PPG data is typically collected from the finger and wrist using different types of light signals. Finger PPG devices use IR and red light, as these are effective in capturing vascular contractions. In contrast, wrist PPG devices utilize green light, which is better at tolerating MA. With the help of the comprehensive UTSA-PPG dataset, we could analyze the performance of different PPG signals in HR/HRV estimation. Fig.~\ref{fig:channel} displays the HR estimation MAPE using different channels in the sit scenario with the UTSA-PPG dataset through the compound method. A multilayer perceptron (MLP) network powered by NAS was employed. During the results, the IR and red lights were collected from finger device, and the green light was collected from wrist device. It shows that, the IR light signals generally outperformed red and green light signals. It also can be found that accuracy varies across different subjects, with the exception of subject 6 and 7, which consistently show lower errors compared to other subjects. This discrepancy could be attributed to personal factors such as age, gender, and other individual characteristics. Additionally, offering multiple PPG channels makes the UTSA-PPG dataset well-suited for developing a comprehensive framework that leverages the advantages of each channel and integrates them to further enhance HR/HRV estimation.

\subsubsection{HR estimation in different scenarios}
\label{sec:compare_scenario}

\begin{figure}
    \centering
    \includegraphics[width=\linewidth]{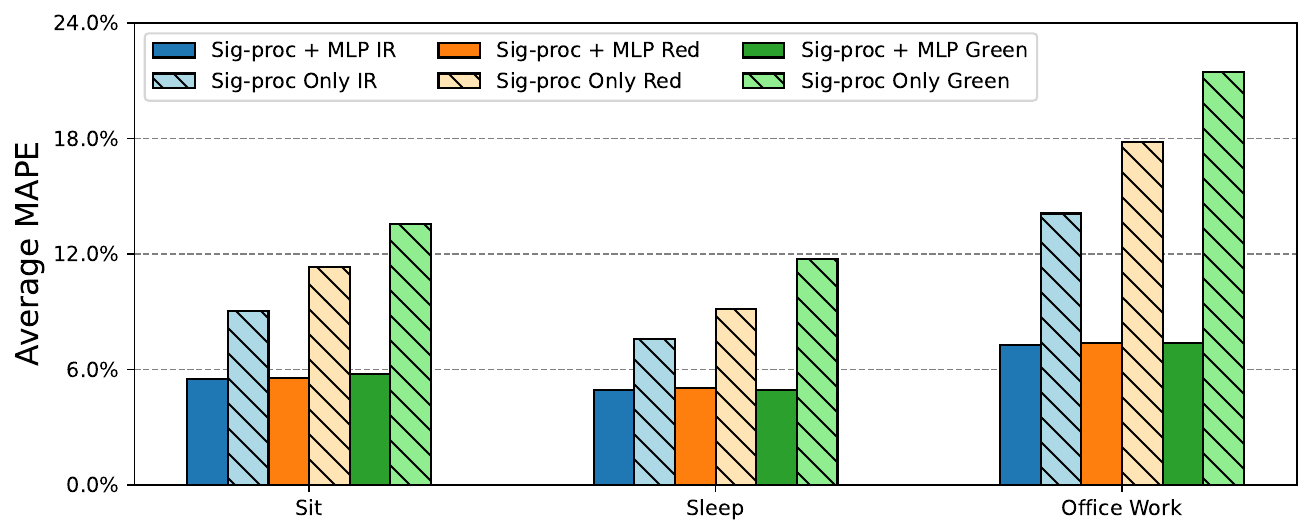}
    \caption{HR estimation in different scenarios.}
    \label{fig:scene}
\end{figure}

Unlike other datasets, the UTSA-PPG dataset includes three distinct activity scenarios that cover a range of movements. Additionally, each scenario features a data session length of at least 30 minutes, supporting long-term monitoring. Therefore, the UTSA-PPG dataset is suitable for studying HR/HRV estimation performance across different scenarios. This is particularly useful for customizing various models tailored to specific scenarios when a single model cannot effectively handle all scenarios. Across all subjects, the average accuracy for different scenarios is shown in Fig.~\ref{fig:scene} to compare the compound method ("Sig-proc + MLP") with the signal processing only method ("Sig-proc Only") for different channels. It can be found that the compound method greatly reduced the error compared to the signal-processing-only method. Additionally, office scenario shows higher errors compared to sit and sleep scenarios for both compound method and the signal-processing-only method. This may be due to higher MA during office work activity. But for the same activity, although the signal-processing-only method provides different errors for different PPG light signals, the compound method provides similar errors for different PPG light signals. This demonstrates the effectiveness of the MLP model which can help the signal processing algorithms to further reduce noise and produce more accurate HR readings.

\subsection{HRV estimation}

To explore the UTSA-PPG dataset's performance on HRV estimation, two common indicators were selected: the Standard Deviation of NN intervals
(SDNN) and the Root Mean Square of Successive Differences (RMSSD), with their equations shown in Equations~\ref{eq_sdnn} and \ref{eq_rmssd}. In both equations, $RR_i$ denotes the $ith$ RR interval, $\overline{RR}$ means the average duration of all $RR$ intervals, and $N$ means the number of intervals.

\begin{equation}
    \label{eq_sdnn}
    \text{SDNN}=\sqrt{\frac{\sum_{i=1}^{N}(RR_i - \overline{RR})^2}{N}}
\end{equation}

\begin{equation}
    \label{eq_rmssd}
    \text{RMSSD}=\sqrt{\frac{\sum_{i=1}^{N-1}(RR_{i+1} - RR_i)^2}{N-1}}
\end{equation}

\subsubsection{HRV estimation with different data history lengths}
\label{sec:compare_data_length}

\begin{figure*}[ht]
    \centering
    \includegraphics[width=0.87\textwidth]{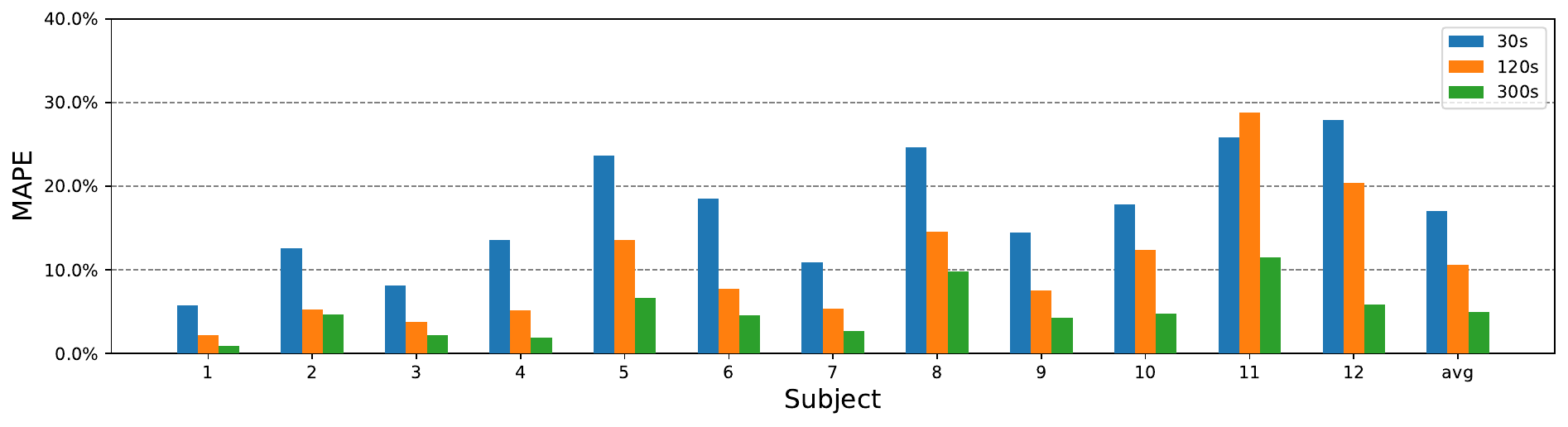}
    \caption{HRV RMSSD estimation with different data history length in sleep scenario.}
    \label{fig:data_history}
\end{figure*}

Another advantage of the UTSA-PPG dataset is its support for long-term monitoring within each scenario, enabling the prediction of various HRV metrics using different data history lengths to meet specific needs. Fig.~\ref{fig:data_history} illustrates HRV estimation performance across different data history lengths in the sleep scenario. The results reveal that the MAPE is significantly lower when using a data history length of 300 seconds. This can be attributed to the fact that long-term monitoring provides more stable HRV estimations, effectively minimizing errors. The stability and accuracy of longer data histories make the UTSA-PPG dataset particularly valuable for applications requiring precise HRV analysis, such as stress monitoring, sleep quality assessment, and cardiovascular research. Furthermore, the ability to adjust data history lengths based on the requirements of specific studies showcases the flexibility of the UTSA-PPG dataset. It allows researchers to customize their models for tasks ranging from real-time HRV monitoring with shorter data histories to more comprehensive evaluations using extended periods of data.

\subsubsection{HRV estimation with different models}
\label{sec:compare_model}

\begin{figure}
    \centering
    \includegraphics[width=\linewidth]{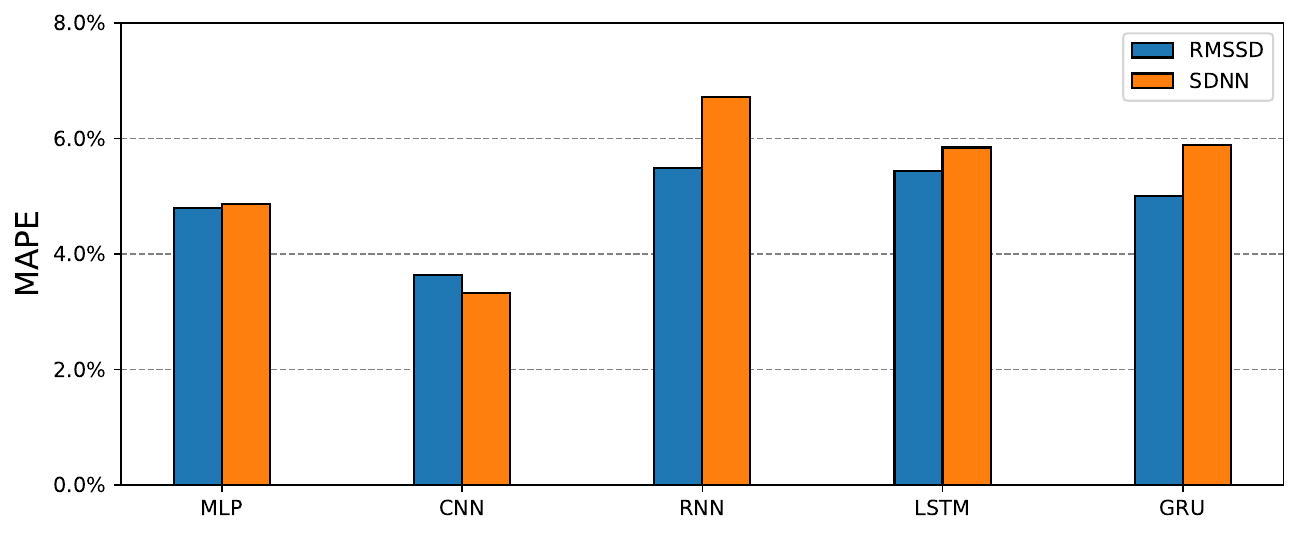}
    \caption{HRV estimations with different models in sit scenario.}
    \label{fig:models}
\end{figure}

The UTSA-PPG dataset also facilitates research on model selection and architecture search for HR/HRV estimation. As illustrated in Fig.~\ref{fig:models}, we evaluated the average error across all subjects in the sit scenario using the compound method~\cite{yuntong2022hrv} with various ML models. In this analysis, HRV was estimated using a data history length of 300s. The results highlight the versatility of the compound method, demonstrating its ability to integrate different ML models while maintaining high accuracy in HR/HRV estimation. Among the tested models, MLP and CNN show superior performance, outperforming RNN, LSTM, and GRU in terms of MAPE. This suggests that, when utilizing compound method, feed-forward and convolutional architectures may be more effective than recurrent models for HR/HRV estimation under the conditions and settings provided in the UTSA-PPG dataset. Such findings emphasize the importance of selecting appropriate model types based on the specific characteristics of the dataset and target application, paving the way for further studies to optimize ML model selection for HR/HRV tasks.
\section{Discussions}

The UTSA-PPG dataset is a comprehensive, long-term dataset that includes multiple activity scenarios, aiming to cover a broad spectrum of daily human activities. However, it has some limitations. While the three designed scenarios encompass most daily activity patterns and are sufficient for studying MA elimination in HR/HRV estimation, the dataset lacks a wider variety of patterned and unpatterned activities, which are essential for evaluating the effects of different activities (long-term and short-term) on HR/HRV estimation. To enhance diversity, we recruited participants with varying skin colors, ages, and genders. However, the dataset lacks patient-specific data, limiting its applicability to HR/HRV estimation research rather than broader statistical analyses. As a result, it is not well-suited for studies that require HR/HRV trend analysis (disease prediction) or distribution-based statistical evaluations. In short, the UTSA-PPG dataset supports robust HR/HRV algorithm evaluation in dynamic, real-world conditions but is less suitable for research focused on predicting diseases based on HR/HRV trends or conducting distribution-based statistical analyses.

\section{Conclusion and Future Work}

In this paper, we conducted an in-depth review of state-of-the-art HR/HRV datasets, examining their strengths and limitations. To address the shortcomings, we developed a custom PPG data acquisition system and collected a comprehensive dataset aimed at advancing research in HR/HRV. By incorporating diverse scenarios, extended session lengths, and multimodal signals, the UTSA-PPG dataset enables robust HR/HRV algorithm evaluation in dynamic, real-world conditions. Supporting both short-term and long-term analyses, as well as facilitating the exploration of multimodal approaches, this dataset serves as a versatile and reliable resource to drive innovation and foster new advancements in HR and HRV estimation. In the future, we plan to carry out the following works using the UTSA-PPG dataset: Incorporate acceleration data into HR/HRV estimation and develop a comprehensive framework capable of accommodating flexible modalities; Investigate methods for reconstructing corrupted PPG data through a self-assessment system for PPG signals; evaluate the performance of state-of-the-art neural network architectures and language models in HR/HRV estimation.

%%%%%%%%%%%%%%%%%%%%%%%%%%%%%%%%%%%%%%%%%%%%%%%%%%%%%%%%%%
\section*{Acknowledgment}
This work was partially supported by the National Science Foundation grants, 2155096, 2215359, and 2306596. The views and conclusions contained herein are those of the authors and should not be interpreted as necessarily representing the official policies or endorsements, either expressed or implied of NSF. The authors would like to thank the anonymous reviewers for their insightful comments.

% \section*{Acknowledgment}

% The preferred spelling of the word ``acknowledgment'' in America is without 
% an ``e'' after the ``g''. Avoid the stilted expression ``one of us (R. B. 
% G.) thanks $\ldots$''. Instead, try ``R. B. G. thanks$\ldots$''. Put sponsor 
% acknowledgments in the unnumbered footnote on the first page.

\bibliographystyle{unsrt}
\bibliography{ref}

\end{document}